\documentclass[11pt]{article}

\usepackage{amsthm,amsmath,bm,amssymb,fullpage,times}

\newtheorem{theorem}{Theorem}

\newtheorem{lemma}[theorem]{Lemma}

\newtheorem{algorithm}{Algorithm}[section]
\newtheorem{remark}[theorem]{Remark}

\newcommand{\ket}[1]{\left|#1\right>}
\newcommand{\bra}[1]{\left<#1\right|}
\newcommand{\C}{{\mathbb{C}}}
\newcommand{\Z}{{\mathbb{Z}}}
\newcommand{\F}{{\mathbb{F}}}
\newcommand{\E}{{\mathbb{E}}}
\newcommand{\R}{{\mathbb{R}}}

\newcommand{\GL}[0]{{\rm GL}}
\newcommand{\onemat}{{\mathbf{1}}}
\newcommand{\zeromat}{{\mathbf{0}}}
\newcommand{\nix}[1]{{}}

\begin{document}

\title{Quantum algorithms to solve the hidden shift problem for 
quadratics and for functions of large Gowers norm}

\author{Martin R{\"o}tteler}

\author{Martin R{\"o}tteler\\
NEC Laboratories America\\
4 Independence Way, Suite 200\\
Princeton, NJ 08540, U.S.A.\\
{\tt mroetteler@nec-labs.com}
}

\maketitle

\begin{abstract}
  Most quantum algorithms that give an exponential speedup over
  classical algorithms exploit the Fourier transform in some way.  In
  Shor's algorithm, sampling from the quantum Fourier spectrum is used
  to discover periodicity of the modular exponentiation function. In a
  generalization of this idea, quantum Fourier sampling can be used to
  discover hidden subgroup structures of some functions much more
  efficiently than it is possible classically.  Another problem for
  which the Fourier transform has been recruited successfully on a
  quantum computer is the hidden shift problem.  Quantum algorithms
  for hidden shift problems usually have a slightly different flavor
  from hidden subgroup algorithms, as they use the Fourier transform
  to perform a {\em correlation} with a given reference function,
  instead of sampling from the Fourier spectrum directly. In this
  paper we show that hidden shifts can be extracted efficiently from
  Boolean functions that are quadratic forms. We also show how to
  identify an unknown quadratic form on $n$ variables using a linear
  number of queries, in contrast to the classical case were this takes
  $\Theta(n^2)$ many queries to a black box. What is more, we show
  that our quantum algorithm is robust in the sense that it can also
  infer the shift if the function is close to a quadratic, where we
  consider a Boolean function to be close to a quadratic if it has a
  large Gowers $U_3$ norm.
\end{abstract}

\section{Introduction}

Fourier analysis has a wide range of applications in computer science
including signal processing, cryptography, Boolean functions, just to
name a few. The fast Fourier transform (FFT) algorithm provides an
efficient way to compute the discrete Fourier transform of length $N$
in time $O(N \log N)$. This is a significant improvement over the
naive $O(N^2)$ implementation and allows to apply Fourier analysis to
correlation problems, to image and audio processing, efficient
decoding of error-correcting codes, data compression, etc. In a more
theoretical context, the Fourier transform over the Boolean
hypercube---also called Walsh-Hadamard transform---is used to study
certain classes of Boolean functions, for instance monotone functions,
functions with constant depth, and functions with variables of high
influence.

In quantum computing, Fourier transforms have turned out to be
extremely successful tools and feature prominently in quantum
algorithms that achieve exponential speedups. The prime examples are
Shor's algorithms for discrete log and factoring \cite{Shor:97}.
Indeed, the quantum computer can sample from the Fourier spectrum on
$N$ points in quantum time $O(\log^2 N)$, a big advantage over the
classical case. Here ``quantum time'' is measured in terms of
elementary quantum gates that are needed to implement the unitary
operation corresponding to the Fourier transform. This possibility of
performing a quantum Fourier transform more efficiently than in the
classical case has a tremendous upside and much of the power of
quantum computing stems from there. This fact has been leveraged for
instance for the solution of the abelian hidden subgroup problem (HSP)
which essentially is solved by sampling from the Fourier spectrum of a
given function \cite{ME:98,BH:97,Kitaev:97}. The hidden subgroup, a
secret property of the function, can then be inferred by a subsequent
classical post-processing.
 
However, the high hopes that Fourier sampling might lead to efficient
quantum algorithms for HSPs over general non-abelian groups, including
cases that would encompass the famous graph isomorphism problem, have
been somewhat dampened recently, as \cite{HMRRS:2006} showed that new
techniques to design highly entangling measurements would be required
in order for the standard approach to succeed.  Perhaps for this
reason, the field of quantum algorithms has seen a shift towards other
algebraic problems such as the algorithm for finding hidden nonlinear
structures \cite{CSV:2007}. The techniques to tackle those problems
are still based on Fourier analysis but have a different flavor than
the HSP.

Classically, besides allowing for {\em sampling} from the spectrum the
importance of the Fourier transform for performing {\em correlation}
tasks cannot be overstated. Therefore, it is very natural to try to
leverage the quantum computer's exponential speedup at computing
Fourier transforms to compute correlations efficiently. It turns out,
however, that this task is an extremely challenging one. First of all,
it can be shown that it is impossible to compute correlations between
two {\em unknown} vectors of data due to requirement for the time
evolution to be unitary and the fact that the correlation between two
inputs is a non-linear map of the inputs.  For some special problems,
however, in which one of the inputs is a fixed, {\em known} vector of
data, correlations can be computed. This question becomes relevant in
particular for hidden shift problems, where correlations can be used
in a particularly fruitful way. These problems ask to identify a
hidden shift provided that access to a function $f(x)$ and a shifted
version $g(x)=f(x+s)$ of the function is given.  Formally, the hidden
shift problem is defined as follows:

\bigskip

\begin{tabular}{rl}
{\bf Given:} & Finite group $G$, finite set $R$, maps $f, g:
G\rightarrow R$.\\ {\bf Promise:} & There exists $s \in G$
such that 
$g(x) = f(x+s)$ for all $x\in G$.\\
{\bf Task:} & Find $s$.
\end{tabular}

\bigskip

The first example of a problem of this kind that was solved on a
quantum computer was $f(x)$ being the Legendre symbol and $s$ being an
unknown element of the cyclic group $\Z_p$ modulo a prime. As shown
in \cite{vDHI:2003}, for the Legendre symbol the hidden shift $s$ can
be found efficiently on a quantum computer. The key observation is
that the Legendre function is an eigenfunction of the Fourier
transform for the cyclic group $\Z_p$.  This fact can be used to
compute a correlation of a shifted Legendre symbol with the Legendre
symbol itself by using the convolution theorem, involving the
application of two discrete Fourier transforms over $\Z_p$.

{\bf Our results.} We present an efficient quantum algorithm to solve
the hidden shift problem for a class of quadratic Boolean functions
for which the associated quadratic form is non-degenerate. Those
functions are special cases of what is known as {\em bent} functions
\cite{Rothaus:76}. An intriguing property of these functions is that,
in absolute values, they have a perfectly flat Fourier spectrum.  In
general, bent functions are those Boolean functions for which the
Hamming distance to the set of all linear Boolean functions is
maximum, where distance is measured by Hamming distance between their
truth tables. A quantum algorithm to solve the hidden shift problem
for bent functions has been studied in \cite{Roetteler:2008}, where
the emphasis is on the richness of different classes of bent functions
for which a hidden shift problem can be defined and solved. In this
paper, in contrast, we restrict ourselves to just one class of bent
functions, namely the quadratic forms, and study a different question:
is it possible to solve the hidden shift problem also in cases where a
given function $f$ is actually not a quadratic form, but close to a
quadratic form? We answer this question in the affirmative, provided
that $f$ is not ``too far'' from a quadratic form, where we measure
closeness by the Gowers norm. We give a quantum algorithm that can
find a hidden shift for functions that are close to quadratics by
using a simple idea: first, we give a quantum algorithm that finds
this quadratic form. Then we solve the hidden shift problem for this
quadratic form by resorting to the hidden shift algorithm for the bent
function case (the case where the corresponding quadratic form is not
of full rank can be taken care without major complications), and
finally we use a test to determine whether the resulting candidate
shift is indeed the correct answer. Overall, we obtain an algorithm
that solves the hidden shift for functions of large Gowers norms using
$O(n)$ queries to the functions. The classical lower bound for such
functions is at least $\Omega(n^2)$ (for the case of perfect
quadratics), but we conjecture that for the case of functions that are
close to quadratics, actually the classical query complexity scales
exponentially.

{\bf Related work.} We already mentioned \cite{Roetteler:2008} which
addressed the hidden shift problem for bent functions and which
constitutes a building block for our algorithm. The hidden shift
problem itself goes back to \cite{vDHI:2003}, in which an algorithm
similar to our Algorithm \ref{alg:standard} was used in order to
correlate a shifted function with a given reference function, thereby
solving a deconvolution problem. The main difference with the present
work is the departure from functions that have perfectly flat Fourier
spectrum.

Our algorithm in Section \ref{sec:quads} to identify the quadratic
function is similar to the methods used in
\cite{CSV:2007,DDW:2008,BCvD:2005} to extract information about
functions that have been encoded into the phases of quantum states.
Related to the considered hidden shift problem is also the work by
Russell and Shparlinski \cite{RS:2004} who considered shift problems
for the case of $\chi(f(x))$, where $f$ is a polynomial on a finite
group $G$ and $\chi$ a character of $G$, a general setup that includes
our scenario.  The two cases for which algorithms were given in
\cite{RS:2004} are the reconstruction of a monic, square-free
polynomial $f \in \F_p[X]$, where $\chi$ is the quadratic character
(Legendre symbol) over $\F_p$ and the reconstruction of a hidden shift
over a finite group $\chi(sx)$, where $\chi$ is the character of a
known irreducible representation of $G$.  The technique used in
\cite{RS:2004} is a generalization to the technique of
\cite{vDHI:2003}. It should be noted that we use a different technique
in our algorithm, namely we combine and entangle two states that are
obtained from querying the function, whereas \cite{RS:2004} has more
the flavor of a ``single register'' algorithm. Another difference is
that our algorithm is time efficient, i.\,e., fully polynomial in the
input size, whereas \cite{RS:2004} is query efficient only.

In a broader context, related to the hidden shift problem is the
problem of unknown shifts, i.\,e., problems in which we are given a
supply of quantum states of the form $\ket{D+s}$, where $s$ is random,
and $D$ has to be identified. Problems of this kind have been studied
by Childs, Vazirani, and Schulman \cite{CSV:2007}, where $D$ is a
sphere of unknown radius, Decker, Draisma, and Wocjan \cite{DDW:2008},
where $D$ is a graph of a function, and Montanaro
\cite{Montanaro:2009}, where $D$ is the set of points of a fixed
Hamming-weight. The latter paper also considers the cases where $D$
hides other Boolean functions such as juntas, a problem that was also
studied in \cite{AS:2007}.  

\section{Fourier analysis of Boolean functions}

First we briefly recall the Fourier representation of a real valued
function $f : \Z_2^n \rightarrow \R$ on the $n$-dimensional Boolean
hypercube. For any subset $S \subseteq [n]=\{1, \ldots, n\}$ there is
a character of $\Z_2^n$ via $\chi_S: x \mapsto (-1)^{S x^t}$, where
$x\in \Z_2^n$ (the transpose is necessary as we assume that all
vectors are row vectors) and $S\in\Z_2^n$ in the natural way.  The inner product of two functions on the
hypercube is defined as $\langle f,g\rangle = \frac{1}{2^n}\sum_x f(x)
g(x) = \E_x(fg)$. The $\chi_S$ are inequivalent character of $\Z_2^n$,
hence they obey the orthogonality relation $\E_x(\chi_S \chi_T) =
\delta_{S,T}$. The Fourier transform of $f$ is a function
$\widehat{f}: \Z_2^n \rightarrow \R$ defined by
\begin{equation}\label{eq:usualDFT}
\widehat{f}(S) = \E_x(f \chi_S) = \frac{1}{2^n} \sum_{x \in \Z_2^n}
\chi_S(x) f(x),
\end{equation}
$\widehat{f}(S)$ is the Fourier coefficient of $f$ at frequency $S$,
the set of all Fourier coefficients is called the Fourier spectrum of
$f$ and we have the representation $f = \sum_S \widehat{f}(S) \chi_S$.
The convolution property is useful for our purposes, namely that
$\widehat{f*g}(S) = \widehat{f}(S) \widehat{g}(S)$ for all $S$ where
the convolution $(f * g)$ of two functions $f$, $g$ is the function
defined as $(f*g)(x) = \frac{1}{2^n} \sum_{y \in \Z_2^n} f(x+y) g(y)$.
In quantum notation the Fourier transform on the Boolean hypercube
differs slightly in terms of the normalization and is given by the
unitary matrix $$H_{2^n} = \frac{1}{\sqrt{2^n}}\sum_{x,y\in \Z_2^n}
(-1)^{xy^t} \ket{x}\bra{y},$$ which is also sometimes called Hadamard
transform \cite{NC:2000}.  Note that the Fourier spectrum defined with
respect to the Hadamard transform which differs from
(\ref{eq:usualDFT}) by a factor of $2^{-n/2}$. It is immediate from
the definition of $H_{2^n}$ that it can be written in terms of a
tensor (Kronecker) product of the Hadamard matrix of size $2\times 2$,
namely $H_{2^n} = (H_2)^{\otimes n}$, a fact which makes this
transform appealing to use on a quantum computer since can be computed
using $O(n)$ elementary operations. 

For Boolean functions $f:\Z_2^n \rightarrow \Z_2$ with values in
$\Z_2$ we tacitly assume that the real valued function corresponding
to $f$ is actually $F : x \mapsto (-1)^{f(x)}$.  The Fourier transform
is then defined with respect to $F$, i.\,.e, we obtain that
\begin{equation}\label{eq:unitaryDFT}
  \widehat{F}(w) =
  \frac{1}{2^n} \sum_{x \in \Z_2^n} (-1)^{wx^t+f(x)},
\end{equation}
where we use $w\in \Z_2^n$ instead of $S\subseteq [n]$ to denote the
frequencies. Other than this notational convention, the Fourier
transform used in (\ref{eq:unitaryDFT}) for Boolean valued functions
and the Fourier transform used in (\ref{eq:usualDFT}) for real valued
functions are the same.  In the paper we will sloppily identify
$\widehat{f} = \widehat{F}$ and it will be clear from the context
which definition has to be used.

We review some basic facts about Boolean quadratic functions. Recall
that any quadratic Boolean function $f$ has the form
$f(x_1,\ldots,x_n) = \sum_{i<j} q_{i,j} x_i x_j + \sum_i \ell_i x_i$
which can be written as $f(x) = x Q x^t + Lx^t$, where $x=(x_1,
\ldots, x_n)\in \Z_2^n$.  Here, $Q\in \F_2^{n\times n}$ is an upper
triangular matrix and $L \in \F_2^{n}$.  Note that since we are
working over the Boolean numbers, we can without loss of generality
assume that the diagonal of $Q$ is zero (otherwise, we can absorb the
terms into $L$). It is useful to consider the associated symplectic
matrix $B = (Q + Q^t)$ with zero diagonal which defines a symplectic
form ${\cal B}(u,v) = u B v^t$.  This form is non-degenerate if and
only if ${\rm rank}(B)=n$.  The coset of $f+R(n,1)$ of the first order
Reed-Muller code is described by the rank of $B$. This follows from
Dickson's theorem \cite{MS:77} which gives a complete classification
of symplectic forms over $\Z_2$:

\begin{theorem}[Dickson \cite{MS:77}]\label{th:Dickson}
  Let $B\in \Z_2^{n \times n}$ be symmetric with zero
  diagonal (such matrices are also called symplectic matrices). Then
  there exists $R \in {\rm GL}(n, \Z_2)$ and $h\in [n/2]$ such that
  $RBR^t = D$, where $D$ is the matrix $(\onemat_{h} \otimes \sigma_x)
  \oplus \zeromat_{n-2h}$ considered as a matrix over $\Z_2$ (where
  $\sigma_x$ is the permutation matrix corresponding to $(1,2)$). In
  particular, the rank of $B$ is always even. Furthermore, under the
  base change given by $R$, the function $f$ becomes the quadratic form
  $ip_h(x_1, \ldots, x_{2h}) + L^\prime(x_1, \ldots, x_{n})$ where we
  used the inner product function $ip_h$ and a linear function
  $L^\prime$.
\end{theorem}

Let $f(x) = x Q x^t + Lx^t$ be a quadratic Boolean function such that
the associated symplectic matrix $B=(Q+Q^t)$ satisfies ${\rm
  rank}(B)=2h=n$. Then the corresponding quadratic form is a so-called
{\em bent function} \cite{Rothaus:76,Dillon:75,MS:77}. In general,
bent functions are characterized as the functions $f$ whose Fourier
coefficients $\widehat{f}(w) = \frac{1}{2^n} \sum_{x \in \Z_2^n}
(-1)^{wx^t+f(x)}$ satisfy $|\widehat{f}(w)| = {2^{-n/2}}$ for all $w
\in \Z_2^n$, i.\,e., the spectrum of $f$ is flat.  It is easy to
see that bent functions can only exist if $n$ is even and that affine
transforms of bent functions are again bent functions.  Indeed, let
$f$ be a bent function, let $A\in \GL(n, \Z_2)$ and $b \in \Z_2^n$,
and define $g(x) := f(xA+b)$. Then also $g(x)$ is a bent function and
$\widehat{g}(w) = (-1)^{-wb} \widehat{f}(w(A^{-1})^t)$ for all $w\in
\Z_2^n$. A very simple, but important observation is that if $f$ is
bent, then this implicitly defines another Boolean function via
$2^{n/2} \widehat{f}(w) =: (-1)^{\widetilde{f}(w)}$. Then this
function $\widetilde{f}$ is again a bent function and called the dual
bent function of $f$. By taking the dual twice we obtain $f$ back:
$\widetilde{\widetilde{f}} = f$.

Theorem \ref{th:Dickson} allows us to define a whole class of bent
functions, namely the Boolean quadratics for which $B=(Q+Q^t)$ has
maximal rank.  It is easy to see that under suitable choice of $Q$, so
instance the inner product function $ ip_n(x_1, \ldots, x_{n}) =
\sum_{i=1}^{n/2} x_{2i-1} x_{2i}$ can be written in this way. Using
affine transformations we can easily produce other bent functions from
the inner product function and Theorem \ref{th:Dickson} also implies
that up to affine transformations the quadratic bent functions are
equivalent to the inner product function. From this argument also
follows that the dual of a quadratic bent function is again a
quadratic bent function, a fact that will be used later on in the
algorithm for the hidden shift problem over quadratic bent functions.

\section{The hidden shift problem for quadratics}\label{sec:quads}

Let $n \geq 1$ and let ${\cal O}$ be an oracle which gives access to
two Boolean functions $f,g : \Z_2^n \rightarrow \Z_2$ such that there
exists $s\in \Z_2^n$ such that $g(x)=f(x+s)$ for all $x\in \Z_2^n$.
The hidden shift problem is to find $s$ by making as few queries to
${\cal O}$ as possible. If $f$ is a bent function, whence also $g$
since it is an affine transform of $f$, then the hidden shift can be
efficiently extracted using the following {\em standard algorithm}.
Recall that Boolean functions are assumed to be computed into the
phase. This is no restriction, as whenever we have a function
implemented as $\ket{x}\ket{0}\mapsto \ket{x}\ket{f(x)}$, we can also
compute $f$ into the phase as $\ket{x} \mapsto (-1)^{f(x)}$ by
applying $f$ to a qubit initialized in
$\frac{1}{\sqrt{2}}(\ket{0}-\ket{1})$.

\begin{algorithm}[Standard algorithm for the hidden shift problem \cite{vDHI:2003}]\label{alg:standard}\ \\ 
\noindent
Input: Boolean functions $f$, $g$ such that $g(x)=f(x+s)$. Output: hidden shift $s$.
\begin{itemize}
\item[(i)] Prepare the initial state
$\ket{0}$. 
\item[(ii)] Apply Fourier transform $H_2^{\otimes n}$ to prepare equal
  distribution of all inputs: $$\frac{1}{\sqrt{2^n}}\sum_{x\in \Z_2^n}
  \ket{x}.$$
\item[(iii)] Compute the shifted function into the
phase to get $$\frac{1}{\sqrt{2^n}}\sum_{x\in \Z_2^n}
(-1)^{f(x+s)}\ket{x}.$$
\item[(iv)] Apply $H_2^{\otimes
  n}$ to get $$\sum_w (-1)^{sw^t} \hat{f}(w) \ket{w} =
\frac{1}{\sqrt{2^n}} \sum_w (-1)^{sw^t} (-1)^{\widetilde{f}(w)}
\ket{w}.$$
\item[(v)] Compute the function $\ket{w} \mapsto
(-1)^{\widetilde{f}(w)}$ into the phase resulting in
$$\frac{1}{\sqrt{2^n}}\sum_w (-1)^{sw^t} \ket{w}.$$
\item[(vi)] Finally, apply another Hadamard transform
  $H_2^{\otimes n}$ to get $\ket{s}$ and measure $s$. 
\end{itemize}
\end{algorithm}

The function $\widetilde{f}$ that has been used in Step (iv) can only
be applied by means of a unitary operation if the Fourier spectrum of
$f$ is flat, in other words if $f$ is a bent function. See also
\cite{Roetteler:2008} for several classes of bent functions to which
this algorithm has been applied. Note that Algorithm
\ref{alg:standard} requires only one query to $g$ and one query to
$\widetilde{f}$. Furthermore, the quantum running time is $O(n)$ and
the algorithm is exact, i.\,e., zero error.  Note that Step (iii) of
Algorithm \ref{alg:standard} assumes that the Fourier transform of $f$
is flat.

There is an intriguing connection between the hidden shift problem for
injective functions $f$, $g$ and the hidden subgroup problem over
semidirect products of the form $A \rtimes \Z_2$ where the action is
given by inversion in $A$ \cite{Kuperberg:2005,FIMSS:2003}. In our
case the functions are not injective, however, it is possible to
exploit the property of being bent to derive another injective
``quantum'' function: $F(x) := \frac{1}{\sqrt{2^n}} \sum_y
(-1)^{f(x+y)} \ket{y}$ (similarly a function $G$ can be derived from $g$).
Now, an instance of an abelian hidden subgroup problem in $\Z_2^n
\rtimes \Z_2$ can be defined via the hiding function $H(x,b)$ that
evaluates to $F(x)$, if $b=0$, and to $G(x)$, if $b=1$. This reduction
leads to an algorithm that is different from Algorithm
\ref{alg:standard}, but also can be used to compute the shift.

Now, we consider a different task: we begin with an arbitrary
quadratic Boolean function (not necessarily bent) $f$, which is given
by an oracle ${\cal O}$. We show that $f$ can be discovered using
$O(n)$ quantum queries to ${\cal O}$, whereas showing a lower bound of
$\Omega(n^2)$ classical queries is straightforward. Recall that
Bernstein and Vazirani \cite{BV:97} solved the case of linear function
$f$. We use quadratic forms $f(x_1,\ldots,x_n) = \sum_{i<j} q_{i,j}
x_i x_j + \sum_i \ell_i x_i$ written as $f(x) = x Q x^t + Lx^t$, where
$x=(x_1, \ldots, x_n)\in \Z_2^n$. Here, $Q\in \Z_2^{n\times n}$ is an
upper triangular matrix and $L \in \Z_2^{n}$.  Using the oracle we can
compute the function into the phase and obtain the state
\begin{equation}\label{eq:quadPhase}
\ket{\psi} = \frac{1}{\sqrt{2^n}}\sum_{x \in \Z_2^n}
(-1)^{x Q x^t+Lx+b} \ket{x}. 
\end{equation}
We will show next, that $Q$ and $L$ can be obtained from a linear
number of copies of $\ket{\psi}$. The method uses two such states at a
time and combines them using the unitary transform defined by
\[
T :
\ket{x,y} \mapsto \frac{1}{\sqrt{2^n}}\sum_{z\in \Z_2} (-1)^{zy^t} \ket{x+y,z}.
\]
Note that $T$ can be implemented efficiently on a quantum computer as
it is just a controlled not between each qubit in the $y$ register as
source to the corresponding qubit in the $x$ register as target,
followed by a Hadamard transform of each qubit in the $y$ register.
The following computation shows that $T$ can be used to extract
information about $Q$ from two copies of $\ket{\psi}$.
\begin{eqnarray*}
T \ket{\psi}\otimes \ket{\psi} & = &
T \left(\frac{1}{2^n} \sum_{x,y} (-1)^{x Q x^t + y Q y^t + L(x+y)^t} \ket{x, y}\right)\\
&=&\frac{1}{\sqrt{2^{3n}}}\sum_{x,y,z} (-1)^{x Q x^t + y Q y^t + L(x+y)^t} (-1)^{z y^t} \ket{x+y, z}\\
&=&
\frac{1}{\sqrt{2^{3n}}}\sum_{x,u,z} (-1)^{u Q u^t + u (Q+Q^t)x^t + Lu^t + z (x+u)^t} \ket{u,z}\\
&=& \frac{1}{\sqrt{2^n}}\sum_{u} (-1)^{u Q^t u^t + Lu^t}\ket{u,u(Q+Q^t)}.
\end{eqnarray*}
Hence this state has the form $\frac{1}{\sqrt{2^n}}\sum_u (-1)^{p(u)} \ket{u,u(Q+Q^t)}$,
where $p$ is the quadratic Boolean function $p(u)=uQ^t u^t + Lu^t$ .

We now describe a direct way to recover $f$ from sampling from these
states. Suppose we sample $k=O(n)$ times, obtaining pairs $(u_i, v_i)$
from this process. The goal is to identify the matrix $Q$. Observe
that learning what $Q$ is equivalent to learning what $M := (Q+Q^t)$
is since $Q$ is an upper triangular matrix with zero diagonal. Now,
arrange the sampled vectors $u_i$ into a matrix $U = (u_1 | \ldots |
u_k)$ and similarly $V = (v_1 | \ldots | v_k)$.  Then $U^t M = V^t$ is
a system of linear equations for each of the $n$ columns of $M$. Since
the matrix $U$ was chosen at random, we obtain that it is invertible
with constant probability, i.\,e., we can find $M$ with constant
probability of success. 

We shall now improve this in order to obtain a method that is more
robust regarding errors in the input state $\ket{\psi}$. Instead of
sampling $k$ times from $T \ket{\psi}^{\otimes 2}$, we consider the
coherent superposition $\ket{\psi}^{\otimes 2k}$ and apply $T^{\otimes
  k}$ to it. The resulting state has the form
\begin{equation}\label{eq:stat}
\sum_{u_1, \ldots, u_k} \varphi(u_1, \ldots, u_k) \ket{u_1, \ldots,
  u_k} \ket{M u_1, \ldots, M u_k},
\end{equation} with certain phases, indicated by
$\varphi$. Next, note that there is an efficient classical algorithm
which on input $U$ and $V$ computes the matrix $M$. We can compute
this algorithm in a reversible fashion and apply to the state (\ref{eq:stat}). The resulting state has constant overlap with a state that is
the superposition of the cases for which the Gauss algorithm
computation was successful (returning $M$) and those cases for which
it was unsuccessful (returning $\perp$): using the shorthand notation
$\bold{u} = (u_1, \ldots, u_k)$, we obtain the state
\[
\left(\sum_{\bold{u} \; {\rm good}} \varphi(\bold{u})\ket{\bold{u}}
\ket{\bold{Mu}}\right)\ket{M} + 
\left(\sum_{\bold{u} \; {\rm bad}} \varphi(\bold{u})\ket{\bold{u}}
\ket{\bold{Mu}}\right)\ket{\perp}.
\] 
Measuring this state will yield $M$ with constant probability. Once
$M$ has been found, we can infer $Q$ and uses this information to
compute it into the phases in equation (\ref{eq:quadPhase}) in order
to cancel the quadratic part out. From the resulting states we can
efficiently determine $L$ from a constant number of subsequent Fourier
samplings.

\paragraph{Relation to learning parity with errors}

We now return to the hidden shift problem. In the following we argue
that the quantum algorithm for finding a shift for quadratic functions
has an advantage over classical attempts to do so, since it can even
handle cases where the function is {\em close} to a quadratic
function. It is easy to see that the shift problem for quadratic
functions themselves can be solved classically in $\Theta(n)$ queries:
the lower bound is a straightforward information-theoretic argument.
For the upper bound we show that from knowledge of the quadratics and
the promise that there is a shift $s$ such that $g(x)=f(x+s)$, we can
determine $s$.  Indeed, it is sufficient to query at points
$(0,\ldots, 0)$, and $e_i$, where $e_i$ denotes the $i$th vector in
the standard basis to get equations of the form $s u_i^t = b_i$, where
$u_i \in \Z_2^n$ and $b_i\in \Z_2$. With constant probability after
$n$ trials the solution is uniquely characterized and can be
efficiently found, e.\,g.,  by Gaussian elimination. The problem with
this approach is that if $f$ and $g$ are not perfect quadratics, the
resulting equations will be
\[
s u_1^t \approx_\varepsilon b_1, \;
s u_2^t \approx_\varepsilon b_2, \; \ldots
\]
where the $\approx_\varepsilon$ symbol means that each equation can be
incorrect with probability $1-\varepsilon$. As it turns out from
Theorem \ref{th:closeQuad} below, we will be able to tolerate noise of
the order $\varepsilon = O(1/n)$. It is perhaps interesting to note that 
similar equations with errors have been studied in learning. The best known algorithm is the
Blum-Kalai-Wasserman sieve \cite{BKW:2003}, running in
subexponential time in $n$, albeit able to tolerate constant error $\varepsilon$. 

We show that the following algorithm for computing an approximating
quadratic form is robust with respect to errors in the input function: 
 
\begin{algorithm} {\bf [Find-Close-Quadratic]} The following 
algorithm takes as input a black-box for a Boolean function $f$. The 
output is a quadratic Boolean function which approximates $f$. 
\begin{itemize}
\item Prepare $2k$ copies of the state $\frac{1}{\sqrt{2^n}}\sum_{x}
  (-1)^{f(x)}\ket{x}$.
\item Group them into pairs of $2$ registers and apply the
  transformation $T$ to each pair.
\item Rearrange the register pairs $[1,2]$, $[3,4]$, \ldots,
  $[2k-1,2k]$ into a list of the form $[1,3,\ldots,k,2,4,\ldots,2k]$.
  Next, apply the reversible Gauss algorithm to the sequence of registers. 
\item Measure the register holding the result of the Gauss algorithm
  computation and obtain $M \in \Z_2^{n\times n}$. Use $M$ to
  uncompute the quadratic phase and extract the linear term via
  Fourier sampling.
\end{itemize}
\end{algorithm}

\begin{theorem}\label{th:closeQuad}
  Let $f,g : \Z_2^n \rightarrow \Z_2$ be Boolean functions, let
  $g = \sum_{i,j} q_{i,j} x_i x_j + \sum_i \ell_i x_i$ be a
  quadratic polynomial, and assume that $|\langle f,g \rangle | >
  (1-\varepsilon)$. Then algorithm running {\bf Find-Close-Quadratic}
  on input $f$ finds the quadratic form corresponding to $g$, and
  thereby $g$ itself with probability $p_{success} \geq c
  (1-n\varepsilon)$, where $c$ is a constant independent of $n$.
\end{theorem}

{\em Proof.} First note that $|\langle f, g \rangle|> (1-\varepsilon)$ 
implies that $f$ and $g$ disagree on at most $\varepsilon 2^n$ of 
the inputs. Hence the two quantum states $\ket{\psi_f} = \frac{1}{\sqrt{2^n}} \sum_x f(x) \ket{x}$ and $\ket{\psi_g} = \frac{1}{\sqrt{2^n}} \sum_x g(x) \ket{x}$ satisfy $|\langle \psi_f | \psi_g \rangle | > (1-\varepsilon)$. 

Next, observe that the algorithm can be seen as application of a
unitary operation $U$. We first study the ``perfect'' case, where we
apply $U$ to the state $\ket{\psi_g^{\otimes k}}$ and then study the
effect of replacing this with the input corresponding to $f$. Notice
that the algorithm can also be seen as a POVM ${\cal M}$ which
consists of rank $1$ projectors $\{E_i : i \in I\}$ such that $\sum_{i
  \in I} E_i = \onemat$. Since the algorithm identifies $M$ with
constant probability, we obtain that the POVM element $E_M$, which
corresponds to the correct answer satisfies $Pr({\rm measure} \; $M$)
= {\rm tr}\left(E_M \ket{\psi_g^{\otimes k}} \bra{\psi_g^{\otimes
      k}}\right) = p_0 \geq \Omega(1)$.

For vectors $v$, $w$ we have that $\|v-w\|_2^2 = 2 - 2|\langle v, w
\rangle|$, we get using $|\langle \psi_f^{\otimes k} | \psi_g^{\otimes
  k} \rangle | > (1-\varepsilon)^k \sim (1-k\varepsilon) +
O(\varepsilon^2)$. For the difference $\ket{\delta} :=
\ket{\psi_f^{\otimes}} - \ket{\psi_g^{\otimes}}$ we therefore get that
$\|\delta\|^2 < 2k\varepsilon$.  Denoting $E_M =
\ket{\varphi}\bra{\varphi}$ with normalized vector $\ket{\varphi}$, we
obtain for the probability of identifying $M$ on input $f$:
\begin{eqnarray*}
{\rm tr}\left(E_M \ket{\psi_f^{\otimes k}}\bra{\psi_f^{\otimes k}}\right) 
&=& 
\langle \psi_g | E_M | \psi_g^{\otimes k} \rangle + 
\langle \delta | E_M | \psi_g^{\otimes k} \rangle + 
\langle \psi_g^{\otimes k} | E_M | \delta \rangle + 
\langle \delta | E_M | \delta \rangle\\
&\geq& p_0 + 2 \langle \delta | \varphi \rangle \langle \varphi | \psi_g^{\otimes k} \rangle + |\langle \delta | \varphi \rangle |^2.
\end{eqnarray*}
By Cauchy-Schwartz, we finally get that $|\langle \delta | \varphi
\rangle| \leq \|\delta\| \|\varphi\| \leq \sqrt{2k\varepsilon}$.
Hence, we obtain for the overall probability of success $p_{success}
\geq p_0 - \sqrt{8 k \varepsilon}$.  
\qed

We give an application of Theorem \ref{th:closeQuad} to the problem of
efficiently finding an approximation of a function of large Gowers
$U_3$ norm in the following section. 

\section{Polynomials and the Gowers norm}\label{app:gowers}

Recall that the Gowers norms measure the extent to which a function
$f: \F^n \to \C$ behaves like a phase polynomial. For $k\geq 1$, the
Gowers norm is defined by
\[
\|f\|_{U^k(\F^n)} := \big(\E_{x,h_1,\ldots,h_k \in \F^n}
  \Delta_{h_1} \ldots \Delta_{h_k} f(x)\big)^{1/2^k},
\]
where $\Delta_h f(x) = f(x+h)-f(x)$ for all $h \in \F^n$. It is
immediate that if $|f(x)|\leq 1$ for all $x$, then
$\|f\|_{U^k(F^n)}\in [0,1]$. Moreover, degree $k$ polynomials are
characterized precisely by the vanishing of $\Delta_{h_1} \ldots
\Delta_{h_k} f(x)$ for all $h_i$. It is furthermore easy to see that
$\|f\|_{U^k(\F^n)} = 1$ if and only if $f$ is a phase polynomial of
degree less than $k$ \cite{GT:2008}.

\begin{theorem}[Inverse theorem for the Gowers $U_3$ norm
  \cite{GT:2008}]
  Let $f: \F^n \rightarrow \C$ be a function that is bounded as
  $|f(x)|\leq 1$ for all $x$. Suppose that the $k$th Gowers norm of
  $f$ satisfies $\|f\|_{U^k(F^n)} \geq 1-\varepsilon$.  Then there
  exists a phase polynomial $g$ of degree less than $k$ such that
  $\|f-g\| = o(1)$. For fixed field $\F$ and degree $k$, the $o(1)$
  term approaches zero as $\varepsilon$ goes to zero.
\end{theorem}

Before we state the algorithm we recall a useful method to compare two
unknown quantum states for equality. This will be useful for a
one-sided test that the output of the algorithm indeed is a valid shift. 

\begin{lemma}[SWAP test \cite{Watrous:2000,BCW+:2001}]\label{lem:swap} Let $\ket{\psi}$,
  $\ket{\varphi}$ be quantum states, and denote by SWAP the quantum
  operation which maps $\ket{\psi}\ket{\varphi} \mapsto
  \ket{\varphi}\ket{\psi}$, and by $\Lambda(SWAP)$ the same operations
  but controlled to a classical bit. Apply $(H_2 \otimes \onemat)
  \Lambda(SWAP) (H_2 \otimes \onemat)$ to the state
  $\ket{0}\ket{\varphi}\ket{\psi}$, measure the first qubit in the
  standard basis to obtain a bit $b$ and return the result (where
  result $b=1$ indicates that the states are different). Then $Pr(b=1)
  = \frac{1}{2} - \frac{1}{2} |\langle \varphi| \psi\rangle |^2$.
\end{lemma} 

Lemma \ref{lem:swap} has many uses in quantum computing, see for
instance \cite{Watrous:2000,BCW+:2001}. Basically, it is useful
whenever given $\ket{\varphi}$ and $\ket{\psi}$ two cases have to be
distinguished: (i) are the two states equal, or (ii) do they have
inner product at most $\delta$. For this case it provides a one-sided
test such that $Pr(b=1)=0$ if $\ket{\psi}=\ket{\varphi}$ and $Pr(b=1)\geq
\frac{1}{2}(1-\delta^2)$ if $\ket{\psi}\not=\ket{\varphi}$ and $|\langle
\varphi| \psi\rangle | \leq \delta$.

\begin{algorithm}\label{alg:shiftedLargeU3} {\bf [Shifted-Large-U3]}
  The following algorithm solves the hidden shift problem for an
  oracle ${\cal O}$ which hides $(f, g)$, where $g(x)=f(x+s)$ for
  $s\in \Z_2^n$ and where $\|f\|_{U_3(\Z_2)} \geq (1-\varepsilon)$.
\begin{enumerate}
\item Solve the hidden quadratic problem for $f$. This gives a
  quadratic $g(x) = xQx^t + Lx^t$. 
\item Compute the dual quadratic function corresponding to the Fourier
  transform of $g$.
\item Solve the hidden shift problem for $f(x)$, $f(x+s)$, and $g$.
  Obtain a candidate $s\in \Z_2^n$.
\item Verify $s$ using the SWAP test.
\end{enumerate}
\end{algorithm}

\begin{theorem}
  Let $f$ be a Boolean function with $\|f\|_{U_3} \geq 1-\varepsilon$.
  Then Algorithm \ref{alg:shiftedLargeU3} ({\bf Shifted-Large-U3})
  solves the hidden shift problem for $f$ with probability
  $p_{success} > c (1-\varepsilon)$, where $c$ is a universal
  constant.
\end{theorem} {\em Proof sketch.} In general the fact that large
Gowers $U_3$ norm implies large correlation with a quadratic follows
from the inverse theorem for Gowers $U_3$ norm
\cite{GT:2008,Sam:2007}. For the special case of the field $\Z_2$ and
the large Gowers norm $(1-\varepsilon)$ we are interested in, we use
\cite{AKK+:2003} to obtain a stronger bound on the correlation with
the quadratics. The claimed result follows from \cite{AKK+:2003} and
the robustness of Algorithm \ref{alg:shiftedLargeU3} against errors in
the input functions. 
\qed

\begin{remark}
  \rm It should be noted that in the form stated, Algorithm
  \ref{alg:shiftedLargeU3} only applies to the case where the rank $h
  = {\rm rk}(Q+Q^t)/2 = n/2$ is maximum, as only this case corresponds
  to bent functions. However, it is easy to see that it can be applied
  in case $h < n/2$ as well. There the matrix $(Q+Q^t)$ has a
  non-trivial kernel, defining a $n-2h$ dimensional linear subspace of
  $\Z_2^n$. In the Fourier transform, the function is supported on an
  affine shift of dual space, i.\,e., the function has $2^{2h}$
  non-zero Fourier coefficients, all of which have the same absolute
  value $2^{-h}$. Now, the hidden shift algorithm can be applied in
  this case too: instead of the dual bent function we compute the
  Boolean function corresponding to the first $2h$ rows of
  $(R^{-1})^t$, where $R$ is as in Theorem \ref{th:Dickson} into the
  phase. This will have the effect of producing a shift $s$ lying in
  an affine space $s + V$ of dimension $n-2h$. For $h< n/2$ the shift
  is no longer uniquely determined, however, we can describe the set
  of all shifts efficiently in that case by giving one shift and
  identifying a basis for $V$.
\end{remark}

\section{Conclusions and open problems}

It is an interesting question is whether the quantum algorithm to find
approximations for functions for large Gowers norms $U_2$ and $U_3$
can be used to find new linear and quadratic tests for Boolean
functions. Furthermore, it would be interesting to study the tradeoff
between number of queries and soundness for quantum tests, in analogy
to the results that have been shown in the classical case
\cite{ST:2006}.

\section*{Acknowledgment} 
I would like to thank the anonymous referees for valuable comments,
including the argument presented after Algorithm \ref{alg:standard}
which shows how to relate the hidden shift problem to an abelian
hidden subgroup problem.

\newcommand{\etalchar}[1]{$^{#1}$}


\begin{thebibliography}{HMR{\etalchar{+}}06}

\bibitem[AKK{\etalchar{+}}03]{AKK+:2003}
N.~Alon, T.~Kaufman, M.~Krivelevich, S.~Litsyn, and D.~Ron.
\newblock {Testing low-degree polynomials over ${\rm GF}(2)$}.
\newblock In {\em Proc.~RANDOM-APPROX'03}, volume 2764 of {\em Lecture Notes in
  Computer Science}, pages 188--199, 2003.

\bibitem[AS07]{AS:2007}
A.~Atici and R.~Servedio.
\newblock Quantum algorithms for learning and testing juntas.
\newblock {\em Quantum Information Processing}, 6(5):323--348, 2007.

\bibitem[BCD05]{BCvD:2005}
D.~Bacon, A.~Childs, and {W. van} {D}am.
\newblock From optimal measurement to efficient quantum algorithms for the
  hidden subgroup problem over semidirect product groups.
\newblock In {\em Proc.~FOCS'05}, pages 469--478, 2005.

\bibitem[BH97]{BH:97}
G.~Brassard and P.~{H{\o}yer}.
\newblock {An exact polynomial--time algorithm for Simon's problem}.
\newblock In {\em {Proceedings of Fifth Israeli Symposium on Theory of
  Computing and Systems}}, pages 12--33. ISTCS, IEEE Computer Society Press,
  1997.

\bibitem[BKW03]{BKW:2003}
A.~Blum, A.~Kalai, and H.~Wasserman.
\newblock {Noise-tolerant learning, the parity problem, and the statistical
  query model}.
\newblock {\em Journal of the ACM}, 50(4):506--519, 2003.

\bibitem[{Buh}01]{BCW+:2001}
{Buhrman, H. and Cleve, R. and Watrous, J. and {de Wolf}, R.}
\newblock Quantum fingerprinting.
\newblock {\em Phys. Rev. Letters}, 87:167902 (4 pages), 2001.

\bibitem[BV97]{BV:97}
E.~Bernstein and U.~Vazirani.
\newblock Quantum complexity theory.
\newblock {\em SIAM Journal on Computing}, 26(5):1411--1473, 1997.
\newblock Conference version in Proc. STOC'93, pp.~11--20.

\bibitem[CSV07]{CSV:2007}
A.~Childs, L.~J. Schulman, and U.~Vazirani.
\newblock Quantum algorithms for hidden nonlinear structures.
\newblock In {\em Proc.~FOCS'07}, pages 395--404, 2007.

\bibitem[DDW09]{DDW:2008}
Th. Decker, J.~Draisma, and P.~Wocjan.
\newblock {Efficient quantum algorithm for identifying hidden polynomials}.
\newblock {\em Quantum Information and Computation}, 9:215--230, 2009.

\bibitem[DHI03]{vDHI:2003}
{W. van} Dam, S.~Hallgren, and L.~Ip.
\newblock Quantum algorithms for some hidden shift problems.
\newblock In {\em {Proc.~SODA'03}}, pages 489--498, 2003.

\bibitem[Dil75]{Dillon:75}
J.~Dillon.
\newblock {Elementary Hadamard difference sets}.
\newblock In F.~(et~al.) Hoffman, editor, {\em {Proc. 6th S-E Conf. on
  Combinatorics, Graph Theory, and Computing}}, pages 237--249. Winnipeg
  Utilitas Math., 1975.

\bibitem[FIM{\etalchar{+}}03]{FIMSS:2003}
K.~Friedl, G.~Ivanyos, F.~Magniez, M.~Santha, and P.~Sen.
\newblock {Hidden translation and orbit coset in quantum computing}.
\newblock In {\em Proc.~STOC'03}, pages 1--9, 2003.

\bibitem[GT08]{GT:2008}
B.~Green and T.~Tao.
\newblock {An inverse theorem for the Gowers $U^3(G)$ norm}.
\newblock {\em Proc. Edin. Math. Soc.}, 2008.
\newblock To appear, see also arxiv preprint math.NT/0503014.

\bibitem[HMR{\etalchar{+}}06]{HMRRS:2006}
S.~Hallgren, C.~Moore, M.~R{\"o}tteler, A.~Russell, and P.~Sen.
\newblock Limitations of quantum coset states for graph isomorphism.
\newblock In {\em Proc.~STOC'06}, pages 604--617, 2006.

\bibitem[Kit97]{Kitaev:97}
A.~Yu. Kitaev.
\newblock Quantum computations: algorithms and error correction.
\newblock {\em Russian Math. Surveys}, 52(6):1191--1249, 1997.

\bibitem[Kup05]{Kuperberg:2005}
G.~Kuperberg.
\newblock {A subexponential-time quantum algorithm for the dihedral hidden
  subgroup problem}.
\newblock {\em SIAM Journal on Computing}, 35(1):170--188, 2005.

\bibitem[ME98]{ME:98}
M.~Mosca and A.~Ekert.
\newblock The hidden subgroup problem and eigenvalue estimation on a quantum
  computer.
\newblock In {\em Quantum Computing and Quantum Communications}, volume 1509 of
  {\em Lecture Notes in Computer Science}, pages 174--188. Springer-Verlag,
  1998.

\bibitem[Mon09]{Montanaro:2009}
A.~Montanaro.
\newblock Quantum algorithms for shifted subset problems.
\newblock {\em Quantum Information and Computation}, 9(5\&6):500--512, 2009.

\bibitem[MS77]{MS:77}
F.~J. MacWilliams and N.~J.~A. Sloane.
\newblock {\em The Theory of Error--Correcting Codes}.
\newblock North--Holland, Amsterdam, 1977.

\bibitem[NC00]{NC:2000}
M.~Nielsen and I.~Chuang.
\newblock {\em Quantum Computation and Quantum Information}.
\newblock Cambridge University Press, 2000.

\bibitem[Rot76]{Rothaus:76}
O.~S. Rothaus.
\newblock On ``bent'' functions.
\newblock {\em Journal of Combinatorial Theory, Series A}, 20:300--305, 1976.

\bibitem[R{\"o}t08]{Roetteler:2008}
M.~R{\"o}tteler.
\newblock Quantum algorithms for highly non-linear boolean functions.
\newblock arXiv Preprint 0811.3208, 2008.

\bibitem[RS04]{RS:2004}
A.~Russell and I.~Shparlinski.
\newblock {Classical and quantum function reconstruction via character
  evaluation}.
\newblock {\em Journal of Complexity}, 20(2--3):404--422, 2004.

\bibitem[Sam07]{Sam:2007}
A.~Samorodnitsky.
\newblock Low-degree tests at large distances.
\newblock In {\em Proceedings of the 39th Annual ACM Symposium on Theory of
  Computing (STOC'07)}, pages 506--515, 2007.

\bibitem[Sho97]{Shor:97}
P.~Shor.
\newblock Polynomial-time algorithms for prime factorization and discrete
  logarithms on a quantum computer.
\newblock {\em SIAM Journal on Computing}, 26(5):1484--1509, 1997.

\bibitem[ST06]{ST:2006}
A.~Samorodnitsky and T.~Trevisan.
\newblock {Gowers uniformity, influence of variables, and PCPs}.
\newblock In {\em Proc.~STOC'06}, pages 11--20, 2006.

\bibitem[Wat00]{Watrous:2000}
J.~Watrous.
\newblock Succinct quantum proofs for properties of finite groups.
\newblock In {\em Proc.~FOCS'00}, pages 537--546, 2000.

\end{thebibliography}
\end{document}